\documentclass{jaa}
\usepackage{natbib}
\usepackage{rotating, graphicx}
\usepackage{ulem}
\usepackage{hyperref}

\begin{document}\sloppy
	
	%%paper title
	%%For line breaks \\ can be used within title
	%\title{Title of the paper goes here:\\ Second line}
	\title{\textsc{pyTANSPEC}: A Data Reduction Package for TANSPEC}
	
	%%author names are separated by comma (,)
	%%use \and before the last author name
	%%use a * along with the number separated by comma
	%% for the  author for correspondence
	%%\textsuperscript{number} is used for affiliation
	%%\affilOne, \affilTwo etc., upto \affilTwentyfive is possible
	%%Please note the first letter after \affil is capitalised in the command
	%%

	\author{Supriyo Ghosh\textsuperscript{1,*}, Joe P. Ninan\textsuperscript{1}, Devendra K. Ojha\textsuperscript{1}, and Saurabh Sharma\textsuperscript{2}}
	\affilOne{\textsuperscript{1}Tata Institute of Fundamental Research, Homi Bhabha Road, Colaba, Mumbai 400 005, India.\\}
	\affilTwo{\textsuperscript{2}Aryabhatta Research Institute of Observational Sciences, Manora Peak, Nainital-263001, Uttarakhand, India.}
	
	%%escape two column mode for title, affiliation and abstract
	%%by giving \twocolumn command as shown
	
	\twocolumn[{
		
		\maketitle
		%%include \corres to print the corresponding author Email id
		\corres{supriyoani89@gmail.com}
		
		%%include \msinfo for
		%%manuscript information such as
		%%received, revised and accepted dates
		%%
		\msinfo{...}{...}
		
		%%abstract
		\begin{abstract}
			 The TIFR-ARIES Near Infrared Spectrometer (TANSPEC) instrument provides simultaneous wavelength coverage from 0.55 to 2.5 $\mu$m, mounted on India’s largest ground-based telescope, 3.6-m Devasthal Optical Telescope at Nainital, India. The TANSPEC offers three modes of observations, imaging with various filters, spectroscopy in the low-resolution prism mode with derived R$\sim$ 100 -- 400 and the high-resolution cross-dispersed mode (XD-mode) with derived median R$\sim$ 2750 for a slit of width 0.5 arcsec. In the XD-mode, ten cross-dispersed orders are packed in the 2048 $\times$ 2048 pixels detector to cover the full wavelength regime. As the XD-mode is most utilized as well as for consistent data reduction for all orders and to reduce data reduction time, a dedicated pipeline is at the need. In this paper, we present the code for the TANSPEC XD-mode data reduction, its workflow, input/output files, and a showcase of its implementation on a particular dataset.  This publicly available pipeline \textsc{pyTANSPEC} is fully developed in \textsc{Python} and includes nominal human intervention only for the quality assurance of the reduced data. Two customized configuration files are used to guide the data reduction. The pipeline creates a log file for all the fits files in a given data directory from its header, identifies correct frames (science, continuum and calibration lamps) based upon the user input, offers an option to the user for eyeballing and accepting/removing of the frames, does the cleaning of raw science frames and yields final wavelength calibrated spectra of all orders simultaneously.
		\end{abstract}
		
		%%insert keywords separated by 3 hyphens using \keywords{words}

	}]
	%%close the twocolumn escape here
	
	%%include \doinum{number}for the DOI number in the header
	%%include \volnum{number} for the volume number in the header
	%%include \year{yyyy} for  year of publication in the header
	%%include \pgrange{num--num} page range of article in the header
	%%include \artcitid{num} for the article citation id
	%%include \lp to print last page of the article
	%%include \setcounter{page}{pagenum} for the exact starting page of the article
	
	\doinum{12.3456/s78910-011-012-3}
	\artcitid{\#\#\#\#}
	\volnum{000}
	\year{2021}
	\pgrange{1--10}
	\setcounter{page}{1}
	\lp{1}

	\section{Introduction}
	\begin{figure*}[h!]
		\centering
		\includegraphics[scale=0.5]{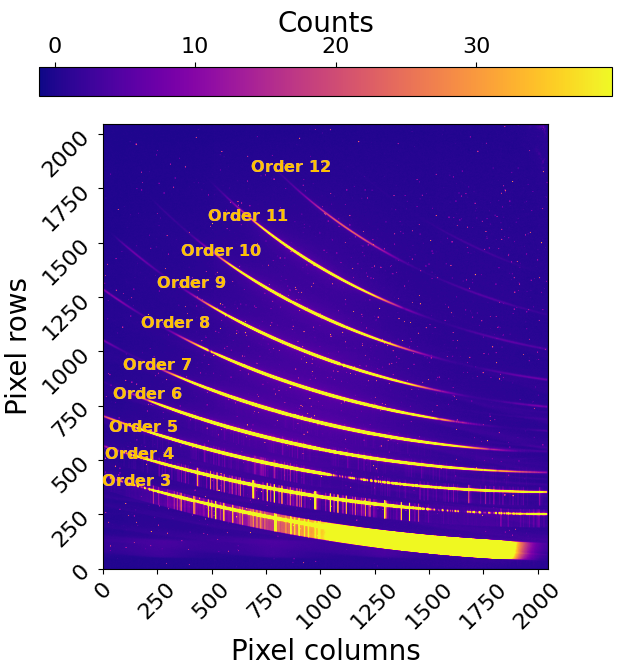}
		\includegraphics[scale=0.5]{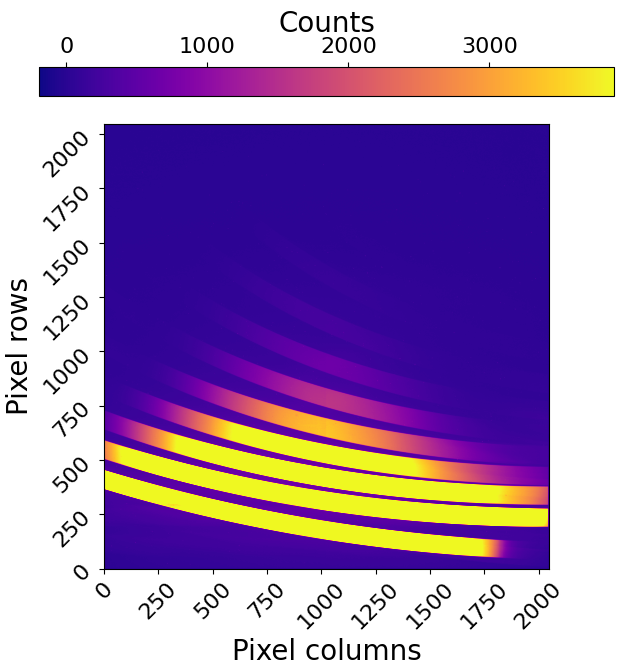}
		\includegraphics[scale=0.5]{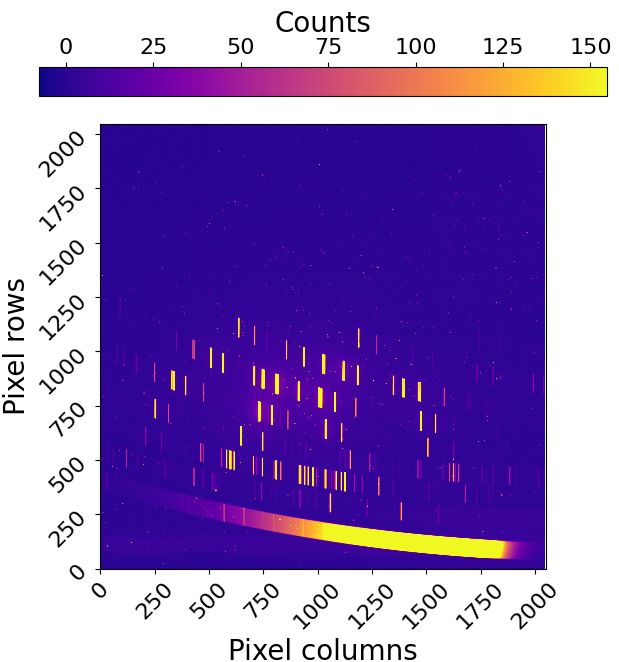}
		\includegraphics[scale=0.5]{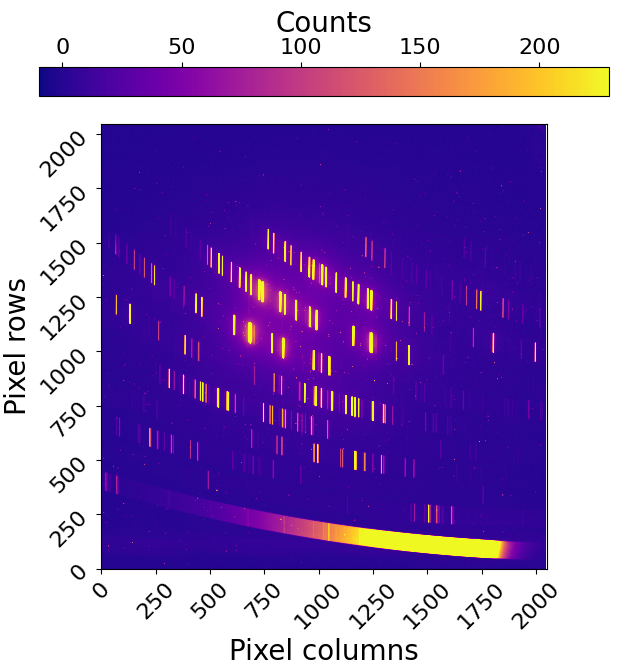}
		\caption{Raw images of an object frame, cont1 lamp, Ar-lamp, and Ne-lamp (from left to right). The colour bar represents counts and its’ scale is from m $-$ 2$\sigma$ to m + 30$\sigma$, where m and $\sigma$ are the sigma clipped median and standard deviation, respectively. All ten orders of the TANSPEC XD-mode, from the XD-order 3 (wavelength$\sim$ 1.9 -- 2.5 $\mu$m) at the bottom to the XD-order 12 (wavelength$\sim$ 0.55 -- 0.61 $\mu$m) at the top, are marked in the object frame.}
		\label{Fig:raw_images}
	\end{figure*}
	%\begin{figure}[h!]
	%	\centering
	%	\includegraphics[scale=0.35]{image_object.png}
	%	\includegraphics[scale=0.35]{image_cont1.png}
	%	\includegraphics[scale=0.35]{image_Ar.png}
	%	\includegraphics[scale=0.35]{image_Ne.png}
	%	\caption{Raw images of an object frame, cont1 lamp, Ne-lamp, and Ar-lamp (from left to right).}
	%	\label{Fig:raw_images}
	%\end{figure}
	The TIFR-ARIES Near Infrared Spectrometer (TANSPEC) is developed in collaboration with Tata Institute of Fundamental Research (TIFR), Mumbai, India; Aryabhatta Research Institute of Observational Sciences (ARIES), Nainital, India and Mauna Kea Infrared LLC (MKIR), Hawaii. The TANSPEC was installed on India’s largest ground-based telescope, 3.6-m Devasthal Optical Telescope (DOT) at Nainital, India on April 02, 2019. It is one of few instruments in the world having simultaneous continuous wavelength coverage from 0.55 to 2.5 $\mu$m with median resolving power, R$\sim$ 2750 in the cross-dispersed mode (XD-mode) and R$\sim$ 100 -- 400 in the low-resolution prism mode (LR-mode) for a slit of width 0.5 arcsec. However, it has slits of different widths ranging from 0.5 to 4.0 arcsec for observations. In addition, it has an independent imaging camera consisting of several broadbands and narrowband filters with a pixel scale of 0.25 arcsec/pixel. Therefore, the TANSPEC provides three modes of observations, XD-mode and LR-mode spectroscopy and imaging-mode using two Teledyne HgCdTe Astronomical Wide Area Infrared Imager (HAWAII)) detectors (H1RG: 1024 $\times$ 1024 pixels for imaging as well as slit-viewing and H2RG: 2048 $\times$ 2048 pixels for spectroscopy). The estimated peak gain value for H1RG is 4.3 $\pm$ 0.2 e$^-$/ADU, while H2RG offers observations in two gain-modes: low-gain and high-gain modes with peak values 4.5 $\pm$ 0.2 e$^-$/ADU and 1.12 $\pm$ 0.03 e$^-$/ADU, respectively (\citealt{2022PASP..134h5002S}). The detectors read out in Sample-Up-The-Ramp (SUTR) mode. In SUTR mode, the voltage in each pixel reads out during the exposure without impacting the ongoing photon collection. Hence, we get a time series of measurements for each pixel, which linearly increases proportionally to the flux falling on the pixel. To generate the final image, the flux in each pixel is calculated by the linear slope fitting along the time axis. The estimated upper limits of the linear regime for H2RG and H1RG detectors are 26000 ADU (57000 ADU for high gain) and 30000 ADU, respectively (\citealt{2022PASP..134h5002S}). We refer to the TANSPEC document\footnote{\url{https://www.aries.res.in/sites/default/files/files/3.6-DOT/Tanspec-Specification-Manual.pdf}} and instrument paper (\citealt{2022PASP..134h5002S}) for additional details.
	%\footnote{\url{https://www.tifr.res.in/}}, \footnote{\url{https://aries.res.in/}}, \footnote{\url{http://mkir.com/}}, \footnote{\url{https://aries.res.in/facilities/astronomical-telescopes/360cm-telescope}}
	
	Among all observation modes, the XD-mode is one of the most efficient ways of studying celestial sources by acquiring spectroscopic data in a broad wavelength range in a single exposure. So, it is the most commonly used mode by the observers. Images in XD-mode observations typically consist of object frames (science target) with corresponding continuum lamps (cont1) and arc lamps (Ar and Ne) for the continuum correction and the wavelength calibration, respectively. The observed frames in the TANSPEC XD-mode, for instance, are shown in Fig.~\ref{Fig:raw_images}. As presented in Fig.~\ref{Fig:raw_images}, the raw images show multiple cross-dispersed orders (10 orders from grating order 3 to 12) in 2048 $\times$ 2048 pixels detector to cover the full-wavelength regime of the TANSPEC. As a result, the data reduction of the XD-mode is a bit more complex and time-consuming than the other two modes (imaging and LR-mode). In addition, the spectral orders are curved and thus, the dispersion axis is not aligned with the detector columns which makes the extraction of XD-mode spectra particularly challenging (\citealt{2004PASP..116..362C}). Hence, a dedicated automated pipeline is at the need for consistent data reduction over the night and for reducing the data reduction time. 
	
	In order to provide a set of semi-automatic scripts for reducing TANSPEC XD-mode data, we have developed a spectral reduction package, \textsc{pyTANSPEC} (version 0.0.1) keeping in mind that the coding language \textsc{Python} is used nowadays by the largest fraction of users among the astronomical community and the development and maintenance of well-known astronomical data reduction software, \textsc{IRAF} is discontinued since 2013. The current version supports data reduction for slits of width 0.5 and 1.0 arcsec and it embarks on slope-fitted images.
	
	In this paper, we illustrate \textsc{pyTANSPEC} and provide an example of spectra reduced with it. We describe the \textsc{pyTANSPEC} package and its workflow in Section 2. %~\ref{sec:package_description}.Section~\ref{sec:Results_and_Discussion}
	Section 3 deals with the reduction and discussion of an A-type star, and the future aspects of \textsc{pyTANSPEC} are given in Section 4. %~\ref{sec:Future_Improvements}.
	Finally, we summarize in Section 5.%~\ref{sec:Summary}.
	
	\section{Description of the Package} \label{sec:package_description}
\begin{figure*}[h!]
		\centering
		\includegraphics[scale=0.9]{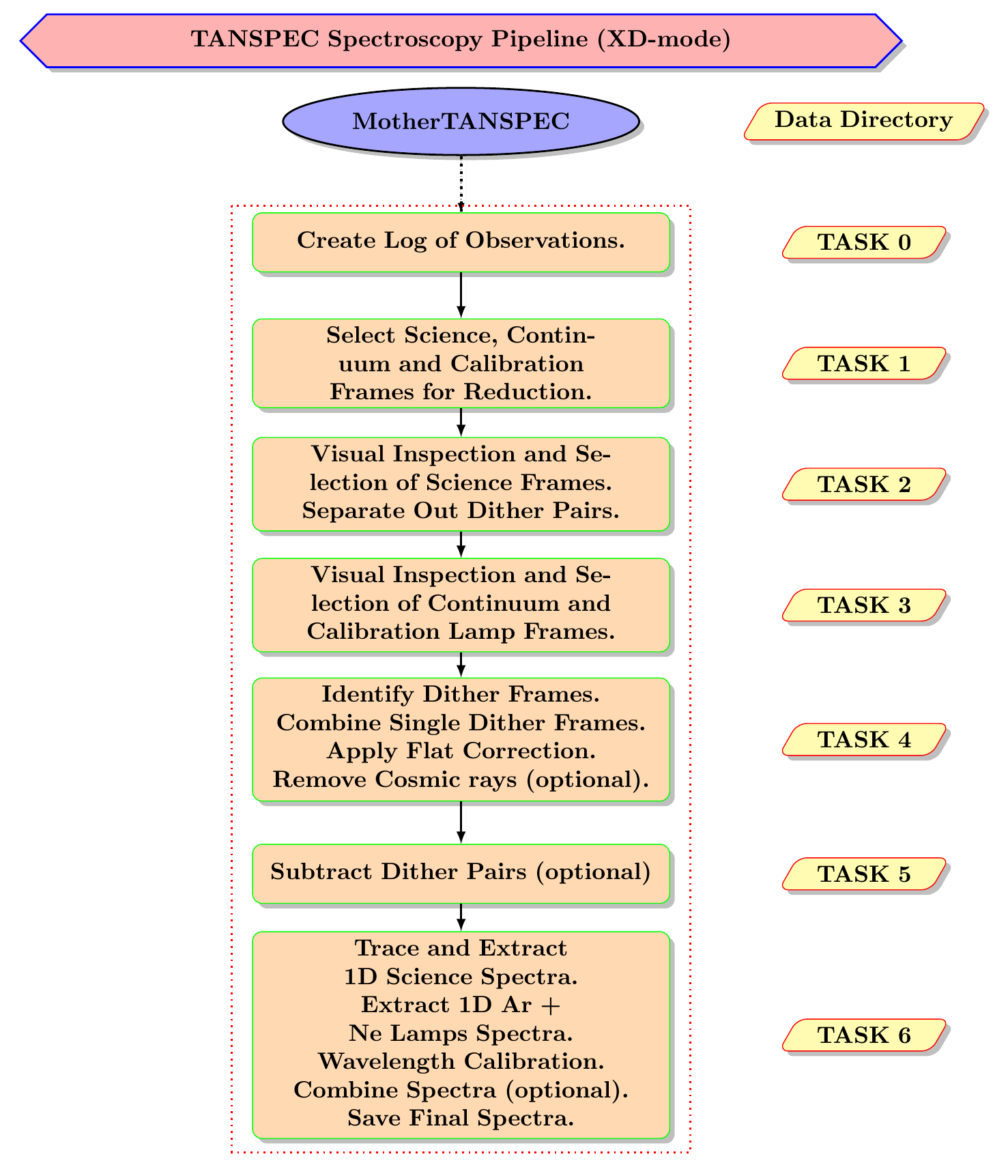}
		\caption{Flow diagram of the TANSPEC XD-mode data reduction pipeline. Users need to create a directory and copy all the data and pipeline-related requirements into it. Users can name the directory according to their whims. Here, we have named it as MotherTANSPEC.}
		\label{Fig:Flow_diagram}
	\end{figure*}

	 The purpose of the pyTANSPEC\_v0.0.1 pipeline is to extract the XD-mode spectra consistently over the night by minimizing human errors that can be introduced during manual reduction as well as reducing the required time of manual data analysis. The pipeline works for object images obtained in either one or two positions of the slit. The data reduction pipeline is solely written in \textsc{Python 3} and has a modular setup. The \textsc{numpy} (\citealt{2020Natur.585..357H}), \textsc{astropy} (\citealt{2013A&A...558A..33A, 2018AJ....156..123A}), \textsc{scipy} (\citealt{2020NatMe..17..261V}), \textsc{scikit-image} (\citealt{2014arXiv1407.6245V}), \textsc{ccdproc} (\citealt{2015ascl.soft10007C}) and \textsc{matplotlib} (\citealt{Hunter:2007}) packages of \textsc{Python} are used for most calculations. Instead of full-scale automation, the pipeline is made semi-automated with minimal human intervention for the quality assurance of the reduced data. The interface of the pipeline has been designed a lot alike the TIFR Near Infrared Spectrometer and Imager (TIRSPEC) pipeline\footnote{\url{https://indiajoe.github.io/TIRSPEC/}} (\citealt{2014JAI.....350006N}) so the users of it can operate the TANSPEC pipeline with ease. Each step of the pipeline can be run independently or simultaneously. The metadata for the pipeline is passed on from one step to the next step via human-readable text files. This helps the users to control the flow of the pipeline. The \textsc{pyTANSPEC} can be downloaded from Github\footnote{\url{https://github.com/astrosupriyo/pyTANSPEC}}. In this paper, we describe the essential steps carried out by the pipeline and for practical implication we refer to the documentation. The flow chart of the pipeline is displayed in Fig.~\ref{Fig:Flow_diagram} and has been described in this section.
	
	\subsection{Preparation}
	All configurations can be done in configuration files. Two configuration files\footnote{\url{https://github.com/astrosupriyo/pyTANSPEC/tree/main/pyTANSPEC/config}}, TANSPECscript.config and spectrum\_extractor\_TANSPEC.config are used to define the required parameters for running the pipeline. There are several parameters, and some of them are redundant. A description of the available parameters can be found in the configuration file itself. The redundant parameters are kept in the configuration files for future developments of the pipeline. TANSPECscript.config is the main configuration file running the pipeline while spectrum\_extractor\_TANSPEC.config is used during the spectrum extraction. In Table~\ref{tab:Description_of_the_input_variables_in_TANSPECscript.conf} we list the most important parameters in the TANSPECscript.config file to be modified to set up the extraction of a new target. However, the default values of parameters can be adopted for the second config file. Thus, two configuration files are used in the pipeline for easing out complexity and making it friendlier to users. After customizing configuration files, the pipeline runs end-to-end to convert raw spectroscopic images into wavelength-calibrated spectra.
	
	\begin{table*}
		%\centering
		\caption{Description of active input variables in TANSPECscript.config}
		\label{tab:Description_of_the_input_variables_in_TANSPECscript.conf}
		
		\resizebox{0.99\textwidth}{!}{
			%\centering
			\begin{tabular}{ll} 
				\hline
				\hline
				Parameter & Short Description \\
				\hline 
				TEXTEDITOR & Required text editor during pipeline run \\
				REMOVE\_COSMIC & To remove cosmic rays \\
				REMOVE\_CONTINUUM\_GRAD &  Remove the illumination pattern in the continuum lamp spectrum using median filter \\
				CONT\_GRAD\_FILT\_SIZE & Specify the median filter box size \\
				Selected\_Window\_Dither\_Find & Specify continuum window for dither pair detection \\
				ApertureWindow & Aperture width (lower and upper limit relative to center) for extracting spectrum \\
				BkgWindows & Background sample regions \\
				WLFitFunc & Function to find the wavelength versus pixels map \\
				SCOMBINE & To average the wavelength calibrated spectra if dithering mode of observation \\ 
				\hline	
				
		\end{tabular}} \\
		  \scriptsize{\textbf{Notes:} While ‘REMOVE\_CONTINUUM\_GRAD’ is made optional in anticipation of our future development of the pipeline,  ‘REMOVE\_COSMIC’ is made optional to opt out of the cosmic ray correction if it does affect the science spectrum. However, for TANSPEC, ‘REMOVE\_CONTINUUM\_GRAD’ should be always ‘True’ as mentioned in the configuration file.}
	\end{table*}
	%\begin{sidewaysfigure}[h!]
	%	\centering
	%	\includegraphics[scale=0.39]{/home/supriyo/All-mother-tanspec/ForPipelineRelease/figure/logofobs.m2.png}
	%	\caption{Contents of created log file. Alternative columns are marked and named in pink and blue colours. It also shows how one can prefix `\#' and discard the particular image (follow the lines in yellow colour). Files are sorted alphabetically and assigned with a number (see the last column). These newly generated numbers should be considered as the file number.}
	%	\label{Fig:logofobs}
	%\end{sidewaysfigure}
	\subsection{Log File Generation}
	In the first step, the pipeline automatically creates an observational log file of the raw frames for a given data directory based on the header information. It lists the name of all files, their corresponding selected header keywords and auto-generated file number in an ascending order started from `0' in a text file inside the given data directory. The auto-generated file number should be used during the pipeline run where required. Creating a log file is an important step for the TANSPEC pipeline to nullify the complexity of running the pipeline because of the user-defined random naming and numbering of files. The users can discard any frame by prepending `\#' to the line of that particular image in the log file if required.
	
	\subsection{Selection, Visualisation and Acceptance of Image Frames}
	In the next three steps, the pipeline yields options to the users for selecting objects as well as continuum and calibration (Ar and Ne lamps) frames, displays all the selected frames and accepts or discards frames based on the user input. These are important steps as the users get another opportunity to discard unwanted frames if they are uncertain which frame to prefix ‘\#’ in the created log file as discussed above as well as to cross-check desired frames for reduction with selected ones.
	
	\subsection{Identification of Dither Frames}
	The conventional way of near-infrared (NIR) observation is to observe the target at two different positions of the slit to counter the sky background. The pipeline is capable of identifying and separating dithered frames. To detect the dithered frames, a small region (80 $\times$ 42 pixels) of a certain continuum window is selected. The continuum window can be customized through the main configuration file by providing the starting and ending pixels of that window along the cross-dispersed axis. Pixel values of cross-dispersed order 5 are provided for this purpose by default. The difference between the starting and ending pixels (80 pixels) serves as the height of the region, while a fixed width of 42 pixels is selected along the dispersion axis centred at the middle of the cross-dispersed axis. Background subtracted median counts (F$_{BS_i}$, where i = 1 ... 80 pixels) are estimated for the defined region along the dispersion axis. We consider only those pixels ($y_{i}$) having counts ($F_{BST_i}$) greater than the threshold value to derive the centre pixel of the dispersion axis for the region following the equation,
	\begin{equation}
		x = \frac{\sum_{i}^{n}F_{BST_i}y_{i}}{\sum_{i}^{n}F_{BS_i}}
	\end{equation}
	The threshold value is defined as the summation of background and 6$\sigma$ counts, where $\sigma$ is the calculated standard deviation using the median absolute deviation of F$_{BS}$. If centres of the raw spectra are greater than 6 pixels apart from each other along the dispersion axis, they are identified as the dither pair by the pipeline.
	
	\subsection{Combining Image Frames}
	The combined images (e.g., science frames at each dither and continuum) are created by the pipeline to improve the SNR. The images are median-combined on a pixel-by-pixel basis. However, the user can also combine images using the sum or average. The combined frames are used for further reduction.
	
		\begin{figure*}[h!]
		\centering
		\includegraphics[scale=0.36]{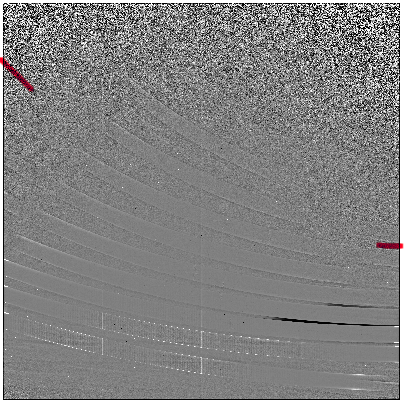}
		\includegraphics[scale=0.36]{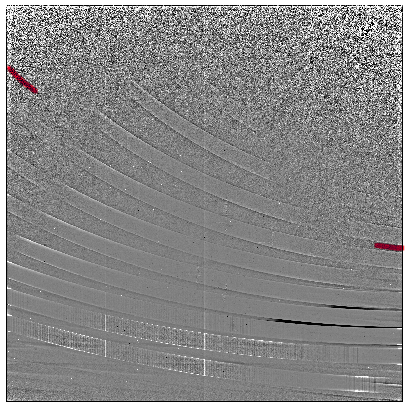}
		\includegraphics[scale=0.36]{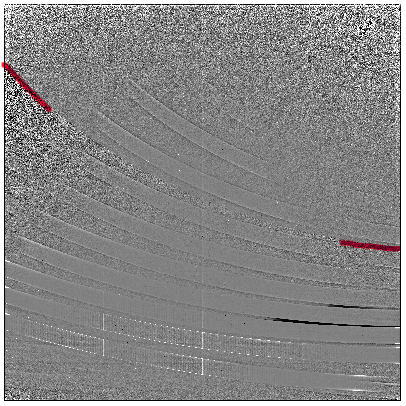}
		\caption{Generation of a final continuum image. Left: Normalised median combined image in each night (F$_{each-night}$), middle: average of normalised median images over several weeks (F$_{avg-several-weeks}$), and right: a final continuum image (F$_{Final}$), where the upper portion of F$_{each-night}$ (an imaginary curved line that connects two marked lines in red colour) is replaced by F$_{avg-several-weeks}$ to form F$_{Final}$, are shown.}
		\label{Fig:master_flat_creation}
	\end{figure*}

	\subsection{Image Processing}
	Image processing includes multiple steps such as correction with dark, bias, and continuum images as well as the removal of cosmic rays and correction with bad-pixel mask for dead pixels, however, processing steps depend on the instrument used for the observations. In the case of TANSPEC, the dark current is small ($\approx$ 0.1 e$^-$/sec in the high-gain setting) and bad pixels are negligible in numbers and the majority of them lie along the edges of the H2RG array (\citealt{2022PASP..134h5002S}), which is used for spectroscopy. Thus, the \textsc{pyTANSPEC} package allows the user to implement flat and cosmic ray corrections. 
	
	\subsubsection{Flat Correction \\ \\}
	The combined continuum frame is smoothened by implementing a median filter. Each pixel of the frame will be replaced by the median value of the surrounding pixels, using a user-defined box (default size: 25 $\times$ 51) around the pixel. The box size of the filter can be customized through the main configuration file. The combined continuum frame is divided by the smoothened version of it to get the normalised continuum image (see the left image in Fig.~\ref{Fig:master_flat_creation}). It can be found that the signal-to-noise ratio (SNR) for higher orders (specifically for orders 10, 11 and 12) of the image is poor. The region lying above the curved line that can be drawn through the marked regions in red colour in Fig.~\ref{Fig:master_flat_creation} is of our concern. We call this imaginary curved line the line-of-control (LoC). To improve the SNR at higher orders, a final continuum image is created (see right one of Fig.~\ref{Fig:master_flat_creation}) where the upper portion of the LoC of the normalised continuum image in each night as described above is replaced by the previously generated normalised master continuum image (see middle one of Fig.~\ref{Fig:master_flat_creation}). The normalised master continuum images come with the \textsc{pyTANSPEC} package and are created by averaging the median filtered continuum data over several weeks of continuum images (a total of 55 images used). They can be found at Github\footnote{\url{https://github.com/astrosupriyo/pyTANSPEC/tree/main/pyTANSPEC/data/CONTLAMPDIR}}. It is found that the SNR above the LoC for the final continuum image is improved by approximately a factor of 4. Using this image, the pipeline does the flat correction.
	
	\subsubsection{Removal of Cosmic Rays \\ \\} 
	Removal of cosmic rays is an optional step in the pipeline. It is performed based on a variation of the Laplacian edge detection algorithm using the \textsc{Python} package \textsc{astrocrappy} (\citealt{2018zndo...1482019M}) which can identify cosmic rays of arbitrary shape and size by the sharpness of their edges. We refer to \citet{2001PASP..113.1420V} for additional details. 
	
	\subsubsection{Subtraction of Dither Pair \\ \\ }
	Dither pair (A and B) is subtracted from each other (i.e. A $-$ B and B $-$ A) to remove the bright NIR sky if the object is observed in two positions of the slit. The subtraction is carried out on a pixel-by-pixel basis. The subtraction of dither pair also removes the dark current and background signal from the telescope, if any. It was made an optional step in the pipeline and depends on the user input during the pipeline run.
	
	\subsection{Spectrum Extraction and Wavelength Calibration}
	\subsubsection{Extracting Science Spectra \\ \\}
	Due to the large curvature in the blue traces of the XD-mode spectrum, it is difficult to trace the blue orders in the 2D image for faint stars. Hence, we took a different approach for the order tracing. We first traced the order positions of a bright star observation and designated that as the template trace. Depending on the position of the star in the slit during the observation, the trace location will shift in the XD direction. Due to atmospheric dispersion as well as spectrograph optics, this will not be a pure linear translation of the trace locations in the XD direction. Hence, in order to match the trace of any given target observations we need to allow the template traces to be non-linearly shifted and stretched in the XD direction. This is done by fitting the cut profiles along the XD direction for all the orders simultaneously using a $2^{nd}$ order polynomial. Even if there is no significant blue light in a particular target's spectrum, the brighter infrared orders will be sufficient to accurately transform the template trace to the location of the stellar spectrum. We have tested the effect of the template fitting on determining the dispersion axis towards the edges and found that our method traces the dispersion axis at the edge approximately identical to the centre. 
		
	Once the trace is matched to the target's spectrum in the 2D image, then the pipeline extracts the spectrum inside user specified aperture window via sum extraction. The sum extraction takes care of the boundary pixel effect by sub-pixel interpolation to sum the light consistently within the exact aperture window size.  The sky background spectrum is also extracted from both sides of the stellar spectrum based on user-specified window sizes. In future, we will add an option for optimal extraction as well which basically scales the spatial profile of an imaged spectrum during its' sum extraction by applying nonuniform pixel weights and reduces the statistical noise to a minimum. For additional details, we refer to \cite{1986PASP...98..609H}.

	\subsubsection{Wavelength Solution \\ \\}
	
	The TANSPEC uses a combination of Ar and Ne lamps for wavelength calibration of all the orders. The exact same aperture window used for the targets' spectrum extraction is used to extract Ar and Ne lamps data via sum extraction. For XD-orders 3--9 (i.e. wavelength range from 0.65 $\mu$m to 2.5 $\mu$m), Ar-lamp data is used, and for the rest (i.e. from 0.55 $\mu$m to 0.75 $\mu$m), Ne-lamp is used. Good lines to use for the fits were manually identified and stored in the pipeline. Without any user input, the pipeline will refit those selected lines from the database with the Gaussian profile model and derive a wavelength dispersion solution. The current version of the pipeline provides Chebyshev and Legendre functions for the wavelength solution which can be customised through the main configuration file. The default value for the wavelength solution is the Legendre polynomial of order 4. Users can verify the goodness of the fit and residuals (derived in terms of radial velocity) from the output directory (we refer to the documentation for details), and accordingly, the order of the fit can be adjusted. However, it is to note that over-fitting can occur on the use of orders greater than 4, especially for spectra of higher XD-orders.
	
	\section{Results and Discussion} \label{sec:Results_and_Discussion}
	We have tested the package with various different data sets for various nights. These include data taken with both slits of width 0.5 and 1.0 arcsec as well as with different gain setups. We will present the reduced spectra of an A-type (HIP 14431) star as an example. It is essential to mention that the reduction procedure as well as the quality of the reduction depend on, in part, the typical approach in which observations were carried out with the TANSPEC.
	
	The star was observed in two dithered positions with a slit of 1.0 arcsec width in a high-gain setting. Two frames (each with an exposure of 180 seconds) were obtained in each dither along with three frames of cont1 and one frame each for Ar and Ne lamps.
	
\begin{figure}[h!]
		\centering
		\includegraphics[scale=0.5]{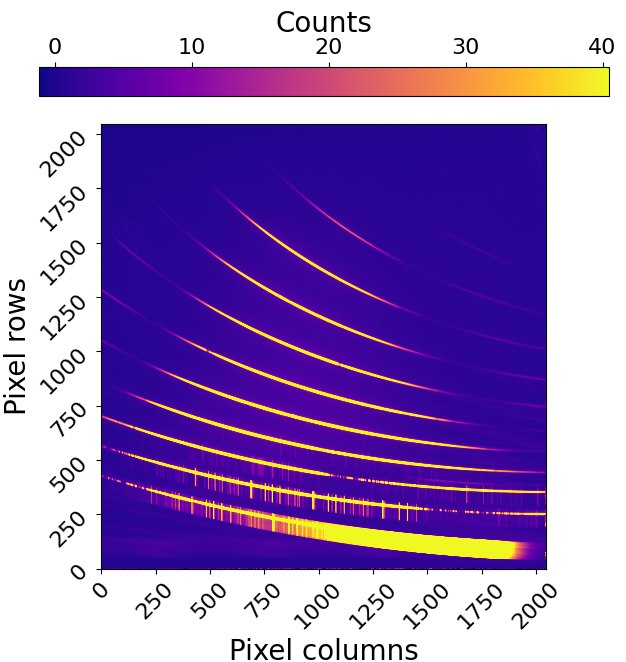}
		\caption{Median combined, flat corrected and cosmic rays removed image for the first dither position. The colour bar represents counts and its’ scale is from m $-$ 2$\sigma$ to m + 30$\sigma$, where m and $\sigma$ are the sigma clipped median and standard deviation, respectively.}
		\label{Fig:flat_corrected_cosmicrays_removed}
	\end{figure}

\begin{figure}[h!]
		\centering
		\includegraphics[scale=0.5]{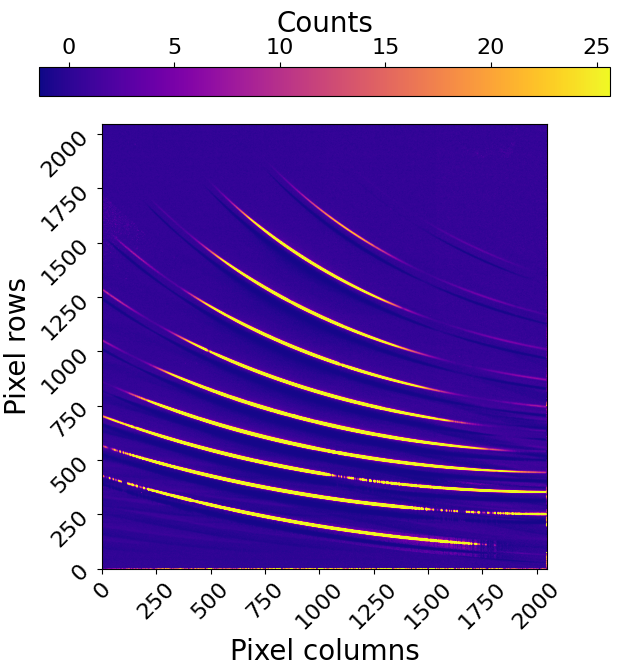}
		\caption{Dither pair subtracted (A $-$ B) image. Colour bar is similar to Fig.~\ref{Fig:flat_corrected_cosmicrays_removed}.}
		\label{Fig:A-B_frame}
	\end{figure}

\begin{figure*}[h!]
		\centering
		\includegraphics[scale=0.65]{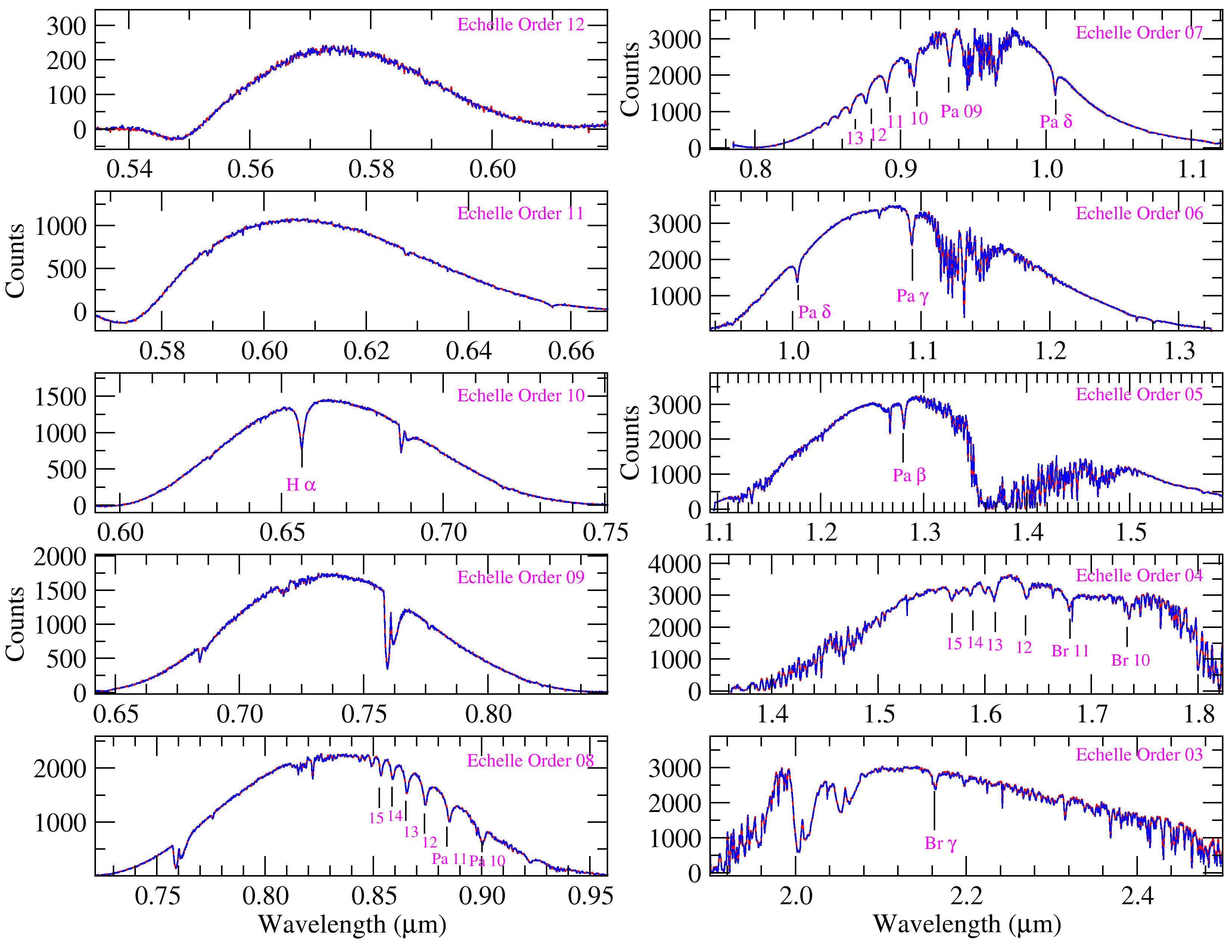}
		\caption{Wavelength calibrated spectra of HIP 14431, an A-type star. Several important spectral lines of the target are also marked. For reduction, we considered two different cases. In the first case, the dither pairs subtraction was performed and extracted spectra are shown in blue colour. For the other case, no dither pairs subtraction was executed and the extracted spectra are overplotted in red colour. In both cases, spectra look similar.}
		\label{Fig:reduced_spectra_for_an_example}
	\end{figure*}

	The data were reduced using the \textsc{pyTANSPEC} pipeline with the parameters described in the manual. Flat correction and cosmic rays removal were implemented to the raw images after the median combined all frames in each dither position. For instance, one median combined, flat corrected and cosmic rays removed image is shown in Fig.~\ref{Fig:flat_corrected_cosmicrays_removed}.  Dither pair was also subtracted to make A $-$ B and B $-$ A frames. After the preparation for the science extraction, the wavelength calibrated spectra of the star were extracted for A $-$ B and B $-$ A frames (see Fig.~\ref{Fig:A-B_frame} for A $-$ B image). Extracted spectra were then averaged out to get the final spectra. The final spectra are displayed in Fig.~\ref{Fig:reduced_spectra_for_an_example}. The full spectral coverage of TANSPEC from 0.55 $\mu$m to 2.5 $\mu$m can be seen. Several spectral features (mainly H-line as it is an A-type star) are identified and marked. Most of the unmarked features are due to terrestrial atmospheric O$_2$ and H$_2$O absorptions. Fig.~\ref{Fig:reduced_spectra_for_an_example} also shows the lower flux (hence lower SNR) in the blue side (i.e. higher orders) in comparison with the red side (i.e. lower orders). It is expected as the grating efficiency falls away significantly towards higher orders from the centre (i.e. zeroth order). 
	
	One of the key aspects of the pipeline is that even when the star is observed in only one position of the slit (which is not followed commonly in NIR observations), the pipeline is capable of extracting spectra effectively. To counter the background counts, the pipeline itself estimates the background using the regions on both sides of the star aperture during the spectrum extraction. As mentioned earlier, the widths of the stars' aperture of x pixels and the background of 2y pixels (y pixels on each side of the star aperture) can be customised through the main configuration file. The estimated backgrounds are then averaged and scaled by the aperture width. Thus, the scaling factor is x/y. The scaled averaged background is subtracted from star flux to remove the same.
	
	For example, we used the same star, HIP 14431 and followed the similar reduction procedures as mentioned above except for the dither pair subtraction (which is an optional step in the pipeline as mentioned earlier). Thus, it can be considered that the star was observed in one position of the slit as we did not subtract dither pair and performed the averaging of extracted spectra of dither pair' images. The spectra is overplotted in Fig.~\ref{Fig:reduced_spectra_for_an_example} in red colour. The extracted spectra for the two cases (with and without the dither pair subtraction) look alike. Thus, the pipeline can extract the spectra efficiently if the star was observed in one position of the slit. The advantage of observing the star in one position of the slit is that it reduces the overhead time during the observation to some extent. In addition, it is important to note that if the star was observed in two dither positions of the slit and the user wants to do dither pair subtraction, the background subtraction has been done twice. The first one is during dither pair subtraction and the second is during the spectrum extraction. We have investigated the possible shift in the dispersion solutionr degrees 4 and 6 for both Legendre and Chebyshev functions on this dataset and found that the possible error in wavelength solution is less than 10 \AA.
		
    Using a Linux machine (Ubuntu 22.04.1 LTS installed), with a 1.60 GHz Intel Core i5 (10th Gen) processor with 8 Gb DDR4 RAM, we found that the pipeline typically takes $\sim$7 minutes to complete all the tasks (i.e. from `TASK 0' to `TASK 6') for this dataset.
	
	\section{Future Improvements} \label{sec:Future_Improvements}
	Several areas for future improvement have been identified. The current version of \textsc{pyTANSPEC} only offers the XD-mode data reduction for slits of width 0.5 and 1.0 arcsec. We plan to include data reduction capability for all available slits in the XD-mode. For that,  normalised master continuum images and lists of previously identified Ar and Ne lines against pixels are only needed to be included for the remaining slits. We expect to release it in the next update. Also, the pipeline does not currently support telluric correction or any form of photometric calibration. We are currently in the process of implementing these features into the pipeline. For spectrum extraction, the pipeline uses a conventional method but the background calculation already performed could be used to implement optimal extraction to maximize SNR. Other areas for improvement are to include the use of continuum lamp (cont2) and scope of reduction for imaging mode as well as LR-mode.   
	
	\section{Summary} \label{sec:Summary}
	The \textsc{pyTANSPEC}, the data reduction package was developed solely in \textsc{Python} for the TANSPEC instrument on 3.6-m DOT to help its users. The pipeline currently supports spectroscopic data reduction for slits of 0.5 and 1.0 arcsec widths for the XD-mode. It is made semi-automated with minimum human interaction for the assurance of quality data reduction. It starts with the generation of a log of observation file that summarises all the information needed about images to run the pipeline. In the following steps, the standard spectroscopic reduction procedures were executed and the pipeline yields wavelength calibrated spectra. Two configuration files that contain several user-defined parameters are used to get control over the pipeline as well as the spectrum reduction. The pipeline typically takes $\sim$ 7 minutes to provide wavelength calibrated spectra of all orders simultaneously from a raw dataset when a Linux machine (Ubuntu 22.04.1 LTS installed) with a 1.60 GHz Intel Core i5 (10th Gen) processor with 8 Gb DDR4 RAM is in use. The pipeline was tested for different data sets over various nights and is performing as per our expectations. Thus, it has been released publicly on March 18, 2022.

	%%Appendix5 × 5 FPS array
	
	\appendix
	
	%\section{An appendix section}
	%Text goes here (Radhakrishnan 1980).
	%\begin{equation}
	%x=a+b+c
	%\end{equation}
	
	%%\section{Another appendix section}
	%Text goes here.
	%\begin{equation}
	%y^2=ax+b+c
	%\end{equation}
	%\vspace{-3em}
	
	%%Use section* for acknowledgements
	\section*{Acknowledgements}
	This work is supported by the Tata Institute of Fundamental Research, Mumbai under the Department of Atomic Energy, Government of India. SG, JPN and DKO acknowledge the support of the Department of Atomic Energy, Government of India, under project Identification No. RTI 4002. We also thank all the members of the IR Astronomy group at TIFR for their support during observations. The authors would like to thank the staff at the 3.6-m DOT, Devasthal and ARIES, for their cooperation during the observation using TANSPEC. SG is thankful to Koshvendra Singh and Arpan Ghosh for running the code before its' public release. We are very much thankful to the referee for their insightful comments, which helped us to improve the paper.
	
	\vspace{-1em}
	
	%%use \balance somewhere in the left column of the last page to balance the two columns in the end page
	
	%%References section
	%\begin{theunbibliography}{}
	%\vspace{-1.5em}
	
	%\bibitem{latexcompanion}
	%Clark D. H., Caswell J. L. 1976, MNRAS, 174, 267
	%\bibitem{latexcompanion}
	%Dickey, J. M., Salpeter, E. E., Terzian, Y. 1978, Astrophys. J. Suppl. Ser., 36, 77
	%\bibitem{latexcompanion}
	%Radhakrishnan, G. C. {\em et al.} 1980, in Evans A., Bode M. F., eds, Non-Solar Gamma Rays (COSPAR), Pergamon Press, %Oxford, p. 163
	%\bibitem{latexcompanion}
	%Starrfield S., Iliadis C., Hix W. R. 2008, in Bode M. F., Evans A., eds, Classical Novae, 2nd edition, Cambridge %University Press, Cambridge, p. 77
	%\bibitem{latexcompanion}
	%Van Loon J. Th. 2008, in Evans A. et al., eds, R S Ophiuchi (2006) and the Recurrent Nova Phenomenon, ASP Conference %Series, Volume 401, p. 90
	%\bibitem{latexcompanion}
	%Zwicky, F. 1957, Morphological Astronomy, Springer-Verlag, Berlin, p. 258
	%

	%\end{theunbibliography}
	%\bibliographystyle{aasjournal}
	\bibliography{reference}

\begin{thebibliography}{}
\expandafter\ifx\csname natexlab\endcsname\relax\def\natexlab#1{#1}\fi

\bibitem[{{Astropy Collaboration} {$et~al$.}(2013){Astropy Collaboration},
  {Robitaille}, {Tollerud}, {Greenfield}, {Droettboom}, {Bray}, {Aldcroft},
  {Davis}, {Ginsburg}, {Price-Whelan}, {Kerzendorf}, {Conley}, {Crighton},
  {Barbary}, {Muna}, {Ferguson}, {Grollier}, {Parikh}, {Nair}, {Unther},
  {Deil}, {Woillez}, {Conseil}, {Kramer}, {Turner}, {Singer}, {Fox}, {Weaver},
  {Zabalza}, {Edwards}, {Azalee Bostroem}, {Burke}, {Casey}, {Crawford},
  {Dencheva}, {Ely}, {Jenness}, {Labrie}, {Lim}, {Pierfederici}, {Pontzen},
  {Ptak}, {Refsdal}, {Servillat}, \& {Streicher}}]{2013A&A...558A..33A}
{Astropy Collaboration}, {Robitaille}, T.~P., {Tollerud}, E.~J., {$et~al$.}
  2013, \aap, 558, A33

\bibitem[{{Astropy Collaboration} {$et~al$.}(2018){Astropy Collaboration},
  {Price-Whelan}, {Sip{\H{o}}cz}, {G{\"u}nther}, {Lim}, {Crawford}, {Conseil},
  {Shupe}, {Craig}, {Dencheva}, {Ginsburg}, {VanderPlas}, {Bradley},
  {P{\'e}rez-Su{\'a}rez}, {de Val-Borro}, {Aldcroft}, {Cruz}, {Robitaille},
  {Tollerud}, {Ardelean}, {Babej}, {Bach}, {Bachetti}, {Bakanov}, {Bamford},
  {Barentsen}, {Barmby}, {Baumbach}, {Berry}, {Biscani}, {Boquien}, {Bostroem},
  {Bouma}, {Brammer}, {Bray}, {Breytenbach}, {Buddelmeijer}, {Burke},
  {Calderone}, {Cano Rodr{\'\i}guez}, {Cara}, {Cardoso}, {Cheedella}, {Copin},
  {Corrales}, {Crichton}, {D'Avella}, {Deil}, {Depagne}, {Dietrich}, {Donath},
  {Droettboom}, {Earl}, {Erben}, {Fabbro}, {Ferreira}, {Finethy}, {Fox},
  {Garrison}, {Gibbons}, {Goldstein}, {Gommers}, {Greco}, {Greenfield},
  {Groener}, {Grollier}, {Hagen}, {Hirst}, {Homeier}, {Horton}, {Hosseinzadeh},
  {Hu}, {Hunkeler}, {Ivezi{\'c}}, {Jain}, {Jenness}, {Kanarek}, {Kendrew},
  {Kern}, {Kerzendorf}, {Khvalko}, {King}, {Kirkby}, {Kulkarni}, {Kumar},
  {Lee}, {Lenz}, {Littlefair}, {Ma}, {Macleod}, {Mastropietro}, {McCully},
  {Montagnac}, {Morris}, {Mueller}, {Mumford}, {Muna}, {Murphy}, {Nelson},
  {Nguyen}, {Ninan}, {N{\"o}the}, {Ogaz}, {Oh}, {Parejko}, {Parley}, {Pascual},
  {Patil}, {Patil}, {Plunkett}, {Prochaska}, {Rastogi}, {Reddy Janga},
  {Sabater}, {Sakurikar}, {Seifert}, {Sherbert}, {Sherwood-Taylor}, {Shih},
  {Sick}, {Silbiger}, {Singanamalla}, {Singer}, {Sladen}, {Sooley},
  {Sornarajah}, {Streicher}, {Teuben}, {Thomas}, {Tremblay}, {Turner},
  {Terr{\'o}n}, {van Kerkwijk}, {de la Vega}, {Watkins}, {Weaver}, {Whitmore},
  {Woillez}, {Zabalza}, \& {Astropy Contributors}}]{2018AJ....156..123A}
{Astropy Collaboration}, {Price-Whelan}, A.~M., {Sip{\H{o}}cz}, B.~M.,
  {$et~al$.} 2018, \aj, 156, 123

\bibitem[{{Craig} {$et~al$.}(2015){Craig}, {Crawford}, {Deil}, {Gomez},
  {G{\"u}nther}, {Heidt}, {Horton}, {Karr}, {Nelson}, {Ninan}, {Pattnaik},
  {Rol}, {Schoenell}, {Seifert}, {Singh}, {Sipocz}, {Stotts}, {Streicher},
  {Tollerud}, {Walker}, \& {ccdproc contributors}}]{2015ascl.soft10007C}
{Craig}, M.~W., {Crawford}, S.~M., {Deil}, C., {$et~al$.} 2015, {ccdproc: CCD
  data reduction software}, ascl:1510.007

\bibitem[{{Cushing} {$et~al$.}(2004){Cushing}, {Vacca}, \&
  {Rayner}}]{2004PASP..116..362C}
{Cushing}, M.~C., {Vacca}, W.~D., \& {Rayner}, J.~T. 2004, \pasp, 116, 362

\bibitem[{{Harris} {$et~al$.}(2020){Harris}, {Millman}, {van der Walt},
  {Gommers}, {Virtanen}, {Cournapeau}, {Wieser}, {Taylor}, {Berg}, {Smith},
  {Kern}, {Picus}, {Hoyer}, {van Kerkwijk}, {Brett}, {Haldane}, {del R{\'\i}o},
  {Wiebe}, {Peterson}, {G{\'e}rard-Marchant}, {Sheppard}, {Reddy}, {Weckesser},
  {Abbasi}, {Gohlke}, \& {Oliphant}}]{2020Natur.585..357H}
{Harris}, C.~R., {Millman}, K.~J., {van der Walt}, S.~J., {$et~al$.} 2020,
  \nat, 585, 357

\bibitem[{{Horne}(1986)}]{1986PASP...98..609H}
{Horne}, K. 1986, \pasp, 98, 609

\bibitem[{Hunter(2007)}]{Hunter:2007}
Hunter, J.~D. 2007, Computing in Science \& Engineering, 9, 90

\bibitem[{{McCully} {$et~al$.}(2018){McCully}, {Crawford}, {Kovacs},
  {Tollerud}, {Betts}, {Bradley}, {Craig}, {Turner}, {Streicher}, {Sipocz},
  {Robitaille}, \& {Deil}}]{2018zndo...1482019M}
{McCully}, C., {Crawford}, S., {Kovacs}, G., {$et~al$.} 2018,
  {Astropy/Astroscrappy: V1.0.5 Zenodo Release}, doi:10.5281/zenodo.1482019

\bibitem[{{Ninan} {$et~al$.}(2014){Ninan}, {Ojha}, {Ghosh}, {D'Costa}, {Naik},
  {Poojary}, {Sandimani}, {Meshram}, {Jadhav}, {Bhagat}, {Gharat}, {Bakalkar},
  {Prabhu}, {Anupama}, \& {Toomey}}]{2014JAI.....350006N}
{Ninan}, J.~P., {Ojha}, D.~K., {Ghosh}, S.~K., {$et~al$.} 2014, Journal of
  Astronomical Instrumentation, 3, 1450006

\bibitem[{{Sharma} {$et~al$.}(2022){Sharma}, {Ojha}, {Ghosh}, {Ninan}, {Ghosh},
  {Ghosh}, {Manoj}, {Naik}, {D'Costa}, {Krishna Reddy}, {Nanjappa}, {Pandey},
  {Sinha}, {Panwar}, {Antony}, {Kaur}, {Sahu}, {Bangia}, {Poojary}, {Jadhav},
  {Bhagat}, {Meshram}, {Shah}, {Rayner}, {Toomey}, {Sandimani}, \& {Pradeep
  R.}}]{2022PASP..134h5002S}
{Sharma}, S., {Ojha}, D.~K., {Ghosh}, A., {$et~al$.} 2022, \pasp, 134, 085002

\bibitem[{{van der Walt} {$et~al$.}(2014){van der Walt}, {Sch{\"o}nberger},
  {Nunez-Iglesias}, {Boulogne}, {Warner}, {Yager}, {Gouillart}, {Yu}, \&
  {scikit-image contributors}}]{2014arXiv1407.6245V}
{van der Walt}, S., {Sch{\"o}nberger}, J.~L., {Nunez-Iglesias}, J., {$et~al$.}
  2014, arXiv e-prints, arXiv:1407.6245

\bibitem[{{van Dokkum}(2001)}]{2001PASP..113.1420V}
{van Dokkum}, P.~G. 2001, \pasp, 113, 1420

\bibitem[{{Virtanen} {$et~al$.}(2020){Virtanen}, {Gommers}, {Oliphant},
  {Haberland}, {Reddy}, {Cournapeau}, {Burovski}, {Peterson}, {Weckesser},
  {Bright}, {van der Walt}, {Brett}, {Wilson}, {Millman}, {Mayorov}, {Nelson},
  {Jones}, {Kern}, {Larson}, {Carey}, {Polat}, {Feng}, {Moore}, {VanderPlas},
  {Laxalde}, {Perktold}, {Cimrman}, {Henriksen}, {Quintero}, {Harris},
  {Archibald}, {Ribeiro}, {Pedregosa}, {van Mulbregt}, \& {SciPy 1. 0
  Contributors}}]{2020NatMe..17..261V}
{Virtanen}, P., {Gommers}, R., {Oliphant}, T.~E., {$et~al$.} 2020, Nature
  Methods, 17, 261

\end{thebibliography}

\end{document}